\documentclass{article}

\title{Components of Antineutrino Emission in Nuclear Reactor}
\author{V. Kopeikin\thanks{kopeykin@polyn.kiae.su}, L. Mikaelyan\thanks{mikaelyan@polyn.kiae.su}, 
V. Sinev\thanks{sinev@polyn.kiae.su}}
\date{Russian Research Centre "Kurchatov Institute", Moscow, Russia}

\begin{document}
\maketitle

Talk given at IV Intern. Conf. NANP'03, Dubna, 23-28 June 2003

\begin{abstract}
New ${\bar{\nu}_e},e$ scattering experiments aimed for sensitive searches of 
the ${\nu}_e$ magnetic moment and projects to explore small mixing angle oscillations at 
reactors call for a better understanding of the reactor antineutrino spectrum. Here we 
consider six components, which contribute to the total ${\bar{\nu}_e}$ spectrum generated in nuclear 
reactor. They are: beta decay of the fission fragments of $^{235}$U, $^{239}$Pu, 
$^{238}$U and $^{241}$Pu, 
decay of beta-emitters produced as a result of neutron capture in $^{238}$U and also due 
to neutron capture in accumulated fission fragments which perturbs the spectrum. 
For antineutrino energies less than 3.5 MeV we tabulate evolution of ${\bar{\nu}_e}$ spectra 
corresponding to each of the four fissile isotopes vs fuel irradiation time and their 
decay after the irradiation is stopped and also estimate relevant uncertainties. 
Small corrections to the ILL spectra are considered.
\end{abstract}

\section*{Introduction}

It has many times been underlined (see e.g. recent reviews [1, 2]) that it is important to 
have an exact knowledge of the reactor antineutrino (${\bar{\nu}_e}$) energy spectrum for planning and 
for analyzing the experiments on neutrino intrinsic properties and on searches for New 
Physics at reactors. 
	
In widely used Pressurized Water Reactors (PWR, VVER in Russian abbreviation) summed U 
and Pu isotope fission rate is $\sim3.1\times10^{19}$/s per 1 GW thermal power. About 
$N_{\nu}\approx 6.7$ ${\bar{\nu}_e}$ are emitted per one fission event. 75\% of them belong to the 
energy range E $<$ 1.80 MeV, below the threshold of the inverse beta decay reaction on proton. 
Quantity $N_{\nu}$ receives contribution from six sources: 
\begin{equation}
N_{\nu} = ^{F}N + ^{U}N + {\delta}^{F}N
\end{equation}
Here $^{F}N\approx 5.5 \ {\bar{\nu}_e}$/fission represents summed contribution from beta decays of 
fission fragments of four fissile isotopes $^{235}$U, $^{239}$Pu, $^{238}$U and $^{241}$Pu, 
undistorted by their interaction with reactor neutrons, $^{U}N\approx 1.2 \ {\bar{\nu}_e}$/fission 
comes from beta decay of $^{239}$U $\Rightarrow \ ^{239}$Np $\Rightarrow \ ^{239}$Pu chain 
produced via neutron radiative capture in $^{238}$U and ${\delta}^{F}N < 0.03 \ {\bar{\nu}_e}$/fission 
originate from neutron capture in accumulated fission fragments and give small but not negligible 
local distortions of the total energy spectrum of the reactor ${\bar{\nu}_e}$.

Plan of this report is as follows:

First, we present a short (and incomplete) overview of a half a century long history, which has led 
to the present understanding of the reactor antineutrinos.

Second, we give new results on the computed evolution of ${\bar{\nu}_e}$ energy spectra corresponding 
to four fissile isotopes vs fuel irradiation time and their decay after the end of the irradiation; 
we compare all available data and estimate relevant uncertainties.

After these data are presented on antineutrinos due to neutron radiative capture in $^{238}$U and in 
accumulated fission fragments.

Finally we consider small corrections to the ILL spectra.

\section{Short history}

Alvarez (unpublished report, 1949 y) did historically the first estimation of the reactor ${\bar{\nu}_e}$
spectrum using conception of fission radiation developed by K. Way and E. Wigner [3]. The next were 
Perkins and King in 1958 y [4]. These studies were stimulated by B. Pontecorvo's proposal to look for 
Cl $\Rightarrow$ Ar transitions near atomic reactor (1946 y) and by famous F. Reines $-$ C. Cowan 
experiments. At that time and many years after it was assumed that the only source of reactor 
antineutrinos is the decay of $^{235}$U fission fragments. 

Kurchatov Institute's Rovno group in 1974-77 yy noticed that in nuclear reactors fission of other 
heavy isotopes produces flux of  comparable to that of $^{235}$U and their energy spectra can be 
quite different from that of $^{235}$U, which, among other effects, would cause time variation of the 
neutrino induced reaction cross sections (burn up effect) [5]. Calculated energy spectrum for 
${\bar{\nu}_e}$ emitted by $^{239}$Pu fission fragments was first published in [6] and for each of 
four fissile isotopes was given in [7]. F. Avignone III et al. in 1980 y were the first to publish 
results on $^{238}$U [8]. B. Davis, P. Vogel et al. in 1979 y [9] also calculated  spectrum for 
$^{239}$Pu and P. Vogel, G. Schenter et al. published results for four fissile isotopes in 1981 y [10]. 
In these publications mainly "high" energy parts ($E > 1-1.5$ MeV) of the antineutrino spectra were 
presented.

Quoted calculations done in 1976-1981 years [6-10] have confirmed the idea that ${\bar{\nu}_e}$ spectra 
associated with fission of different isotopes considerably differ one from another. Thus, relative to 
${\bar{\nu}_e}$ spectrum generated in the decay of $^{235}$U fission fragments, fission of $^{238}$U 
gives much harder spectrum while $^{239}$Pu fission produces ${\bar{\nu}_e}$ of lower energies. 
The absolute values of each of the four spectra was established, however, with large uncertainties 
associated with poor knowledge of the decay schemes of short lived fission fragments, which 
significantly contribute to the hard part of the ${\bar{\nu}_e}$ spectra.

Accurate knowledge of the ${\bar{\nu}_e}$ spectra came from experiments in which relevant 
beta-electron spectra were measured. Electrons and ${\bar{\nu}_e}$ come from the same beta decay 
process and are closely related. This simple idea is used in the conversion method in which 
${\bar{\nu}_e}$ spectrum can be reconstructed (at least for not too low energies) if the spectrum 
of fission electrons is known. This idea was first proposed and used for $^{235}$U by C. Muehlhause 
and S. Oleksa in 1957 [11] and by F. Reines group in 1959 y [12]. Later the correlation between 
calculated electron and ${\bar{\nu}_e}$ fission spectra was analyzed in 1979-1981 yy in already quoted 
papers [9, 10], by K. Schreckenbach et al. [13] in 1981 y and by the Rovno group in 1982 y [14]. First 
experiment in which $^{235}$U and $^{239}$Pu fission electron spectra have been measured and their 
considerable difference observed was performed by Rovno group in 1980-1981 yy [15]. The best 
${\bar{\nu}_e}$ spectra for $^{235}$U, $^{239}$Pu and $^{241}$Pu thermal fission at the neutrino 
energies $E \ge 2.0$ MeV have been found by the ILL group in a number of experiments carried out 
in 1981-1989 yy [16]; they accurately measured relevant electron spectra and effectively modified the 
conversion procedure. These ILL spectra above 2 MeV are commonly used now. For $^{238}$U is used 
spectrum calculated in [10].

Voloshin, Vysotskii and L. Okun' [17], Akhmedov [18], Vogel and Engel [19] in 1986-1989 yy stimulated 
new efforts to search for anomalous magnetic moment of the neutrino in ${\bar{\nu}_e},e$ scattering 
experiments at reactors and further studies of the components of reactor ${\bar{\nu}_e}$ spectrum. 
The Moscow MEPhI group in 1989 y calculated time-evolution of ${\bar{\nu}_e}$ spectra emitted by U 
and Pu fission fragments [20]. 

The fifth component of the spectrum $^{U}N$, which originates from $^{238}$U(n,${\gamma}$) reaction, 
was "discovered" only in 1996 y [21]. The last component ${\delta}^{F}N$, which comes from reactor 
neutron capture in fission fragments was estimated in [7] and considered in [21, 22].

\section{Main results}

\subsection{Fission antineutrinos from four fissile isotopes}
	
For each of four fissile isotope we calculate the time evolution of the ${\bar{\nu}_e}$ spectrum during fuel 
irradiation time $t_{on}$ and its decay as a function of time $t_{off}$ after the end of irradiation. 
Calculations involve summation of all beta branches of 571 fission fragments. For fragments yields 
we use data compiled in [23]. For the decay schemes is used information accumulated in our laboratory 
during past 25 years. Our code evaluates the spectra in the energy range 0-10 MeV (200 points per 
1 MeV) for $t_{on}$ and $t_{off}$ intervals from 0.2 hour to infinity.

Time evolution of the four spectra for ${\bar{\nu}_e}$ energy below 3.5 MeV is presented in Tables 1 
and 2. One can see that at 3.5 MeV full saturation is achieved already in $\sim$1 day time after 
the beginning of fission process, in 2-3 MeV energy range $\sim$ 3\% increase takes place at 
long irradiation times while low energies do not reach equilibrium even in 2 years. (In typical 
PWR reactors 2 year is the average fuel irradiation time at the end of the operational run). 
In Table 3 one can see that at $t_{on}$ = 2 y 50\% of fission ${\bar{\nu}_e}$'s are emitted below 
(1.2-1.3) MeV and $\sim$ 30\% of them have energy higher than 2.0 MeV.

\begin{table}[htb]
\caption{Calculated ${\bar{\nu}_e}$-spectra (1/(MeV$\cdot$fission) for $^{235}$U and $^{239}$Pu vs 
irradiation time $t_{on}$}
\vspace{3pt}
\begin{tabular}{c|c|c|c|c||c|c|c|c}
\hline
Isotope & \multicolumn{4}{|c||}{$^{235}$U} & \multicolumn{4}{c}{$^{239}$Pu} \\
\hline
E,MeV & 1 d & 30 d & 100 d & 2 y & 1 d & 30 d & 100 d & 2 y \\
\hline
0.05 & 0.102 & 0.216 & 0.300 & 0.426 & 0.165 & 0.309 & 0.397 & 0.502\\
0.1 & 0.226 & 0.608 & 0.897 & 1.326 & 0.229 & 0.720 & 1.019 & 1.373\\
0.2 & 0.718 & 1.719 & 2.007 & 2.322 & 0.722 & 2.013 & 2.402 & 2.710\\
0.3 & 1.129 & 2.029 & 2.316 & 2.637 & 1.043 & 1.989 & 2.217 & 2.446\\
0.4 & 1.587 & 2.184 & 2.353 & 2.413 & 1.475 & 2.141 & 2.284 & 2.331\\
0.5 & 1.866 & 2.395 & 2.496 & 2.543 & 1.753 & 2.363 & 2.442 & 2.475\\
0.6 & 1.740 & 2.277 & 2.366 & 2.397 & 1.739 & 2.278 & 2.343 & 2.362\\
0.7 & 1.847 & 2.366 & 2.459 & 2.495 & 1.854 & 2.336 & 2.398 & 2.420\\
0.8 & 1.868 & 2.386 & 2.486 & 2.527 & 1.886 & 2.360 & 2.426 & 2.451\\
0.9 & 1.873 & 2.355 & 2.450 & 2.493 & 1.885 & 2.321 & 2.382 & 2.409\\
1.0 & 1.812 & 2.137 & 2.203 & 2.247 & 1.798 & 2.093 & 2.132 & 2.160\\
1.2 & 1.702 & 1.929 & 1.988 & 2.033 & 1.580 & 1.776 & 1.809 & 1.840\\
1.4 & 1.541 & 1.621 & 1.661 & 1.702 & 1.399 & 1.461 & 1.482 & 1.513\\
1.6 & 1.472 & 1.515 & 1.522 & 1.542 & 1.316 & 1.355 & 1.362 & 1.386\\
1.8 & 1.378 & 1.407 & 1.412 & 1.432 & 1.215 & 1.240 & 1.245 & 1.270\\
2.0 & 1.241 & 1.257 & 1.262 & 1.282 & 1.082 & 1.095 & 1.101 & 1.125\\
2.25 & 1.054 & 1.064 & 1.068 & 1.086 & 0.909 & 0.916 & 0.921 &0.944\\
2.5 & 0.887 & 0.895 & 0.898 & 0.912 & 0.754 & 0.759 & 0.763 & 0.782\\
2.75 & 0.768 & 0.772 & 0.775 & 0.785 & 0.647 & 0.650 & 0.653 &0.668\\
3.0 & 0.650 & 0.651 & 0.651 & 0.652 & 0.538 & 0.539 & 0.540 & 0.546\\
3.25 & 0.553 & 0.554 & 0.554 & 0.554 & 0.445 & 0.445 & 0.446 &0.450\\
3.5 & 0.452 & 0.452 & 0.452 & 0.452 & 0.355 & 0.355 & 0.356 & 0.358\\
\hline
\end{tabular} 
\end{table}

Fuel continues to emit ${\bar{\nu}_e}$ after the irradiation is stopped (Table 4). In the softest part 
of the spectrum (50-500 keV) the residual ${\bar{\nu}_e}$ emission rate is at a level of $\sim$50-5\% 
during the first month and does not completely vanish at $t_{off}$ = 1 year.

To what extend are reliable calculated ${\bar{\nu}_e}$ spectra in the energy range E $<$ 3 MeV? 
Here, in contrast with the E $>$ 3 MeV region, contribution of well-established beta emitters 
amounts to 85-90\%. We estimate that relative uncertainties here do not exceed 5-6\% (68\% C.L.). 
This estimate is confirmed by comparison of present results with spectra calculated earlier by Vogel 
and Engel [19], by MEPhI group [20] and with the ILL conversion spectra [16] (Fig.1). 

\begin{table}[htb]
\caption{Calculated ${\bar{\nu}_e}$-spectra (1/(MeV$\cdot$fission) for $^{238}$U and $^{241}$Pu vs 
irradiation time $t_{on}$}
\vspace{3pt}
\begin{tabular}{c|c|c|c|c||c|c|c|c}
\hline
Isotope & \multicolumn{4}{|c||}{$^{238}$U} & \multicolumn{4}{c}{$^{241}$Pu} \\
\hline
E,MeV & 1 d & 30 d & 100 d & 2 y & 1 d & 30 d & 100 d & 2 y \\
\hline
0.05 & 0.164 & 0.302 & 0.390 & 0.503 & 0.192 & 0.328 & 0.407 & 0.502\\
0.1  & 0.247 & 0.715 & 1.016 & 1.397 & 0.235 & 0.695 & 0.965 & 1.286\\
0.2  & 0.782 & 2.008 & 2.386 & 2.723 & 0.742 & 1.956 & 2.330 & 2.643\\
0.3  & 1.177 & 2.089 & 2.334 & 2.596 & 1.103 & 2.008 & 2.219 & 2.448\\
0.4  & 1.668 & 2.298 & 2.448 & 2.499 & 1.571 & 2.216 & 2.353 & 2.397\\
0.5  & 1.984 & 2.558 & 2.644 & 2.681 & 1.874 & 2.473 & 2.549 & 2.580\\
0.6  & 1.937 & 2.471 & 2.544 & 2.567 & 1.911 & 2.435 & 2.497 & 2.516\\
0.7  & 2.077 & 2.578 & 2.653 & 2.680 & 2.046 & 2.536 & 2.598 & 2.620\\
0.8  & 2.126 & 2.621 & 2.701 & 2.732 & 2.087 & 2.568 & 2.634 & 2.659 \\
0.9  & 2.156 & 2.608 & 2.681 & 2.714 & 2.098 & 2.537 & 2.597 & 2.624\\
1.0  & 2.113 & 2.420 & 2.470 & 2.504 & 2.036 & 2.331 & 2.366 & 2.395\\
1.2  & 1.954 & 2.162 & 2.205 & 2.242 & 1.812 & 1.998 & 2.028 & 2.061\\
1.4  & 1.810 & 1.880 & 1.908 & 1.943 & 1.645 & 1.701 & 1.720 & 1.754\\
1.6  & 1.752 & 1.793 & 1.800 & 1.821 & 1.568 & 1.603 & 1.612 & 1.642\\
1.8  & 1.659 & 1.686 & 1.691 & 1.713 & 1.456 & 1.479 & 1.486 & 1.517\\
2.0  & 1.514 & 1.528 & 1.533 & 1.555 & 1.318 & 1.330 & 1.337 & 1.368\\
2.25 & 1.332 & 1.341 & 1.345 & 1.365 & 1.138 & 1.145 & 1.151 & 1.179\\
2.5  & 1.158 & 1.164 & 1.168 & 1.184 & 0.964 & 0.969 & 0.974 & 0.998\\
2.75 & 1.028 & 1.032 & 1.035 & 1.047 & 0.839 & 0.842 & 0.846 & 0.865\\
3.0  & 0.895 & 0.896 & 0.897 & 0.900 & 0.712 & 0.713 & 0.715 & 0.724\\
3.25 & 0.775 & 0.776 & 0.776 & 0.779 & 0.598 & 0.599 & 0.600 & 0.606\\
3.5  & 0.653 & 0.654 & 0.654 & 0.655 & 0.491 & 0.491 & 0.492 & 0.495\\
\hline
\end{tabular}
\end{table}

\begin{table}[htb]
\caption{Fraction $N(E)/N_{tot}$ of antineutrinos emitted in the energy 
intervals (0-$E$) for fuel irradiation time $t_{on}$=2 y$^{*}$}
\vspace{3pt}
\begin{tabular}{c|c|c|c|c}
\hline
E, MeV & $^{235}$U & $^{239}$Pu & $^{238}$U & $^{241}$Pu\\
\hline
0.1 & 9.94(-3) & 1.64(-2) & 1.10(-2) & 1.55(-2)\\
0.2 & 4.73(-2) & 6.01(-2) & 4.48(-2) & 5.15(-2)\\
0.3 & 8.84(-2) & 0.105    & 8.04(-2) & 9.04(-2)\\
0.5 & 0.179 & 0.201 & 0.158 & 0.174\\
0.75 & 0.290 & 0.320 & 0.257 & 0.285\\ 
1.0 & 0.399 & 0.436 & 0.357 &  0.394\\
1.5 & 0.570 & 0.606 & 0.517 &  0.560\\
2.0 & 0.700 & 0.734 & 0.646 &  0.692\\
2.5 & 0.798 & 0.827 & 0.749 &  0.792\\
3.0 & 0.869 & 0.893 & 0.827 &  0.865\\
4.0 & 0.951 & 0.965 & 0.926 &  0.951\\
5.0 & 0.983 & 0.990 & 0.971 &  0.984\\
\hline
$N_{tot}$, ${\bar{\nu}_e}$/fission & 5.585 & 5.091 & 6.688 & 5.897\\
\hline
\end{tabular} \\
\vspace{5pt}
$^{*}$ Absolute values of $N_{tot}$ for $t_{on}$=2 years are presented in the last row
\end{table}

\begin{table}[htb]
\caption{Residual ${\bar{\nu}_e}$-emission: Ratios of the current $^{235}$U and $^{239}$Pu fission 
antineutrino spectra vs time after the end of fuel irradiation time $t_{off}$ to that at the 
end of irradiation period $t_{on}$ = 2 years}
\vspace{3pt}
\begin{tabular}{c|c|c|c|c||c|c|c|c}
\hline
Isotope & \multicolumn{4}{|c||}{$^{235}$U} & \multicolumn{4}{c}{$^{239}$Pu} \\
\hline
E,MeV & 1 d & 10 d & 30 d & 1 y & 1 d & 10 d & 30 d & 1 y \\
\hline
0.05 & 0.762 & 0.592 & 0.495 & 6.2(-2) & 0.672 & 0.487 & 0.386 & 4.4(-2)\\
0.1 & 0.830 & 0.651 & 0.543 & 6.6(-2) & 0.834 & 0.606 & 0.478 & 5.1(-2)\\
0.2 & 0.691 & 0.395 & 0.262 & 6.2(-2) & 0.734 & 0.403 & 0.259 & 4.2(-2)\\
0.3 & 0.572 & 0.358 & 0.232 & 4.5(-2) & 0.574 & 0.315 & 0.188 & 3.4(-2)\\
0.4 & 0.343 & 0.180 & 9.5(-2) & 8.3(-3) & 0.367 & 0.163 & 8.2(-2) & 6.6(-3)\\
0.5 & 0.266 & 0.120 & 5.9(-2) & 9.2(-3) & 0.292 & 0.104 & 4.6(-2) & 7.4(-3)\\
0.75 & 0.261 & 0.113 & 5.5(-2) & 2.9(-3) & 0.232 & 8.4(-2) & 3.7(-2) & 3.1(-3)\\
1.0 & 0.194 & 8.1(-2) & 4.9(-2) & 4.2(-3) & 0.168 & 5.6(-2) & 3.1(-2) & 4.7(-3)\\
1.25 & 0.130 & 7.5(-2) & 5.2(-2) & 5.9(-3) & 0.106 & 5.4(-2) & 3.6(-2) & 7.4(-3)\\
1.5 & 6.8(-2) & 4.1(-2) & 3.3(-2) & 7.6(-3) & 6.1(-2) & 3.6(-2) & 2.9(-2) & 1.0(-2)\\
1.75 & 3.8(-2) & 2.0(-2) & 1.8(-2) & 8.4(-3) & 4.3(-2) & 2.6(-2) & 2.4(-2) & 1.2(-2)\\
2.0 & 3.2(-2) & 2.2(-2) & 2.0(-2) & 9.2(-3) & 3.8(-2) & 2.9(-2) & 2.7(-2) & 1.3(-2)\\
2.25 & 3.0(-2) & 2.2(-2) & 2.1(-2) & 9.6(-3) & 3.7(-2) & 3.1(-2) & 3.0(-2) & 1.5(-2)\\
2.5 & 2.8(-2) & 2.1(-2) & 2.0(-2) & 8.8(-3) & 3.6(-2) & 3.1(-2) & 2.9(-2) & 1.4(-2)\\
2.75 & 2.2(-2) & 1.8(-2) & 1.7(-2) & 7.5(-3) & 3.2(-2) & 2.8(-2) & 2.7(-2) & 1.3(-2)\\
3.0 & 2.0(-3) & 1.1(-3) & 1.1(-3) & 5.8(-4) & 1.6(-2) & 1.5(-2) & 1.4(-2) & 7.5(-3)\\
3.25 & 1.6(-3) & 8.7(-4) & 8.4(-4) & 4.5(-4) & 1.3(-2) & 1.2(-2) & 1.1(-2) & 6.0(-3)\\
3.5 & 1.2(-3) & 6.0(-4) & 5.8(-4) & 3.1(-4) & 8.9(-3) & 8.2(-3) & 7.9(-3) & 4.2(-3)\\
\hline
\end{tabular}
\end{table}

\subsection{Antineutrinos from neutron capture in $^{238}$U and in fission fragments}

Nuclear fuel in PWR reactors contains 95-97\% of $^{238}$U. $^{238}$U absorbs $\sim$0.6 neutron per 
fission via (n,${\gamma}$) reaction: $^{238}$U + $n \quad \Rightarrow$  $^{239}$U($E_{max}$ = 
1.26 MeV) $\Rightarrow$ $^{239}$Np($E_{max}$ = 0.71 MeV) $\Rightarrow$ $^{239}$Pu. This process 
contributes $^{U}N \sim$ 20\% to the total  flux. The quantity $^{U}N$ is 
practically constant 
over reactor run and is known with an uncertainty of 5\%. Note, that in reactors with fuel 
elements of natural uranium ${\bar{\nu}_e}$ production rate in the channel considered is 
$\sim$1.5 times higher.

Neutron capture in fission fragments can either increase or decrease intensity depending on energy 
of ${\bar{\nu}_e}$ (Fig. 2). The term ${\delta}^{F}N$ (Eq.1) is negative for energies below 0.9 MeV and positive 
for higher energies; it slowly changes along the reactor run and its contribution to the total 
flux of ${\bar{\nu}_e}$ does not exceed 0.3\%. The negative part of ${\delta}^{F}N$ originates mainly from intensive 
absorption of neutrons in the fragment $^{135}$Xe ($T_{1/2}$ = 9.1 h, $E_{max}$ = 0.91 MeV), which 
is produced in reactor at a rate of $\sim$0.07/fission. Due to very high cross section of 
(n,${\gamma}$) reaction (a few million barn) the majority of $^{135}$Xe nuclei absorb neutrons 
before they decay. 

Neutron interactions with reactor construction materials have been found to contribute less than 
0.3\% to the total flux and have been neglected at this stage of study.

\subsection{Small corrections to the ILL spectra}

The ILL spectra have been obtained after 1-1.5 day exposure time. Their uncertainties are estimated 
as 2.5\%. These spectra do not contain time dependent contributions due to decay of long-lived 
fission fragments (see Tables 1, 2) and due to additional radiation associated with neutron capture 
(Fig. 2), which affect part of the ${\bar{\nu}_e}$ spectra above 1.80 MeV, the threshold of the inverse beta 
decay reaction on proton. We mention here this point because it may appear to be of some importance 
in searches of very small mixing angle oscillations at reactors, which have been discussed at 
this Conference.

\section*{Conclusions}
	
	In this report we have tried to give a short story of the long process of developments in understanding reactor antineutrino source and to outline present status of the problem. New efforts to improve the accuracy may be needed in the future. So far, however, we do not feel challenges coming from current or planned experiment, which could stimulate such efforts.

\section*{Aknowledgements}

This work is partially funded by grants 1246.2003.2 (Support of leading scientific schools) 
and 03-02-16055 (RFBR).

\section*{References}

1. C. Bemporad, G. Gratta and P. Vogel, Rev. Mod. Phys. 74, 297, (2002).\\
2. L. A. Mikaelyan, Yad. Fiz. 65, 1206 (2002) [Phys. At. Nucl. 65, 1173, (2002)].\\
3. Way K., Wigner E.P., Phys. Rev. 73, 1318, (1948).\\
4. R. King and J. Perkins, Phys. Rev. 112, 963, (1958).\\
5. A. Borovoi, L. Mikaelyan, A. Rumyancev, On neutrino control of nuclear reactors, Kurchatov 
Institute report No 12-1453, Moscow, 1974, unpublished;\\
L. Mikaelyan, in Proc. Int. Conf. NEUTRINO'77, Baksan, 1977 (Nauka, Moscow, 1978) Vol. 2, p. 383).\\
6. A. A. Borovoi, Yu. L. Dobrynin and V. I. Kopeikin, Yad. Fiz. 25, 264, (1977). 
[Sov. J. Nucl. Phys. 25, 144, (1977)].\\
7. V. I. Kopeikin, Yad. Fiz. 32, 1507 (1980) [Sov. J. Nucl. Phys. 32, 780, (1980)].\\
8. F. T. Avignone III and Z. D. Greenwood, Phys. Rev. C 22, 594, (1980).\\
9. B. R. Davis, P. Vogel, F. M. Mann and R.E. Schenter, Phys. Rev. C 19, 2259, (1979).\\
10. P. Vogel, G. K. Schenter, F. M. Mann, R. E. Schenter, Phys. Rev. C 24, 1543, (1981).\\
11. C. O. Muehlhause, S. Oleksa, Phys. Rev. 105, 1332, (1957).\\
12. R. E. Carter, F. Reines, J. J. Wagner, M. E. Wyman, Phys. Rev. 113, 280, (1959).\\
13. K. Schreckenbach, H. R. Faust, F. von Feilitzsch, A. A. Hahn, K. Haverkamp, J. L. Vuilleumier, 
Phys. Lett. B 99, 251, (1981).\\
14. A. A. Borovoi, V. I. Kopeikin, L. A. Mikaelyan, S. V. Tolokonnikov, Yad. Fiz. 36, 400, (1982) 
[Sov. J. Nucl. Phys. 36, 232, (1982)].\\
15. A. A. Borovoi, Yu. V. Klimov, V. I. Kopeikin, Preprint IAE-3465/2, 1981, Yad. Fiz. 37, 1345, (1983). 
[Sov. J. Nucl. Phys. 37, 801, (1983)].\\
16. F. v. Feilitzsch, A. A. Hahn, K. Schreckenbach, Phys. Lett. B 118, 162, (1982); K. Schreckenbach, 
G. Colvin, W. Gelletly, F. v. Feilitzsch, Phys. Lett. B 160, 325, (1985); A. A. Hahn, K. Schreckenbach, 
W. Gelletly, F. v. Feilitzsch, G. Colvin G., B. Krusche, Phys. Lett. B 218, 365, (1989).\\
17. M. Voloshin, M. Vysotskii and L. Okun', Zh. Exsp. Teor. Fiz., 91, 754, (1986);
      [Sov. Phys. JETP 64, 446, (1986)].\\
18. E. Akhmedov, Yad. Fiz. 48, 599, (1988) [Sov. J. Nuc. Phys. 48, 382, (1988)].\\
19. P. Vogel and J. Engel, Phys. Rev. D39, 3378 (1989).\\
20. V. G. Aleksankin, S. V. Rodichev, P. M. Rubtchov, P. M. Rudjanski and F. E. Chukreev, Beta- and 
antineutrino radiation from radioactive nuclei, Moskow-1989. \\
21. A. M. Bakalyarov, V. I. Kopeikin, L. A. Mikaelyan, Yad. Fiz. 59, 1225 (1996) [Phys. At. Nucl. 59, 1171 
(1996)].\\
22. V. I. Kopeikin, L. A. Mikaelyan, V. V. Sinev, Yad. Fiz. 60, 230 (1997) [Phys. At. Nucl. 60, 172 (1997)].\\
23. T. R. England, B. F. Rider, Evaluation and Compilation of Fission Product Yields, 1993 // LA-UR-94-3106. ENDF-349. 
Los Alamos National Laboratory, October, 1994.\\



\end{document}